\documentclass[a4paper,onecolumn,11pt,eqsecnum,accepted=2024-05-23]{quantumarticle}
\pdfoutput=1
\usepackage[utf8]{inputenc}
\usepackage[english]{babel}
\usepackage[T1]{fontenc}
\usepackage{amsmath,amssymb}
\usepackage{soul}
\usepackage{hyperref}
\usepackage[numbers,compress]{natbib}
\usepackage{tikz}
\usepackage{lipsum}

\usepackage{graphicx,amsmath,color,bm,rotating,dsfont,stackrel,ulem,braket,newtxtext}

\newcommand{\eq}[1]{\begin{equation}\begin{aligned}#1\end{aligned}\end{equation}}
\newcommand{\iu}{\text{i}}
\newcommand{\eu}{\text{e}}
\newcommand{\ha}{\hat{a}}
\newcommand{\had}{\hat{a}^\dagger\vphantom{a}}

\newcommand{\Tr}{\mathop{\mathrm{Tr}} \nolimits}

\newcommand{\im}{\mathop{\mathrm{Im}} \nolimits}

\newcommand{\vac}{\ket{\mathrm{vac}}}
\newcommand{\normord}[1]{:\mathrel{#1}:}
\newcommand{\expct}[1]{\left\langle#1\right\rangle}

\begin{document}

\title{Covariant operator bases for continuous variables}

\author{Aaron Z. Goldberg}
\affiliation{National Research Council of Canada, 100 Sussex Drive, Ottawa, Ontario K1N 5A2, Canada}
\affiliation{Department of Physics, University of Ottawa,  25 Templeton Street, Ottawa, Ontario K1N 6N5, Canada}

\author{Andrei~B.~Klimov}
\affiliation{Departamento de Física, Universidad de Guadalajara, 
44420~Guadalajara, Jalisco, Mexico}

\author{Gerd Leuchs}
\affiliation{Max-Planck-Institut für die Physik des Lichts, 91058 Erlangen, Germany}

\author{Luis~L.~S\'anchez-Soto}
\affiliation{Max-Planck-Institut für die Physik des Lichts, 91058 Erlangen, Germany}
\affiliation{Departamento de Óptica, Facultad de Física, Universidad Complutense, 28040 Madrid, Spain}
  
\begin{abstract}
Coherent-state representations are a standard tool to deal with continuous-variable systems, as they allow one to efficiently visualize quantum states in phase space. Here, we work out an alternative basis consisting of monomials on the basic observables, with the crucial property of behaving well under symplectic transformations. This basis is the analogue of the irreducible tensors widely used in the context of SU(2) symmetry. Given the density matrix of a state, the expansion coefficients in that basis constitute the multipoles, which describe the state in a canonically covariant form that is both concise and explicit. We use these quantities to assess properties such as quantumness or Gaussianity and to furnish direct connections between tomographic measurements and quasiprobability distribution reconstructions.   
\end{abstract}

\maketitle

\section{Introduction}

The notion of observable plays a central role in quantum physics~\cite{Kraus:1983aa}. The term was first used by Heisenberg~\cite{Heisenberg:1925aa}  (\textit{beobachtbare Gr\"{o}\ss e}) to refer to quantities involved in physical measurements and thus having an operational meaning. They give us information about the state of a physical system and may be predicted by the theory. According to the conventional formulation, observables are represented by selfadjoint operators acting on the Hilbert space associated with the system~\cite{Reed:1975aa,Bonneau:2001aa}. 

Given an abstract observable, one has to find its practical implementation. For discrete degrees of freedom, the associated Hilbert space is finite dimensional and the observable is then represented by a matrix whose explicit form depends on the basis. Choosing this basis such that it possesses specific properties can be tricky~\cite{Schwinger:1960a,Wootters:1989aa,Renes:2004aa,Durt:2010aa}. Especially, when the system has an intrinsic symmetry, the basis should have the suitable transformation properties under the action of that symmetry. This idea is the rationale behind the construction of irreducible tensorial sets~\cite{Fano:1959ly}, which are crucial for the description of rotationally invariant systems~\cite{Blum:1981aa} and can be generalized to other invariances~\cite{Kasperkovitz:2003aa}. In this way, the quantum state is represented by its associated multipoles, which are precisely the moments of the generators arranged in a manifestly invariant form.

Things get more complicated in the continuous-variable setting, when the Hilbert space has infinite dimensions. The paradigmatic example is that of a single bosonic mode, where the Weyl-Heisenberg group emerges as a hallmark of noncommutativity~\cite{Binz:2008oq}. As Fock and coherent states are frequently regarded as the most and least quantum states, respectively, they are typically used as bases in quantum optics.  Coherent states constitute an overcomplete basis which is at the realm of the phase-space formulation of quantum theory~\cite{Tatarskii:1983uq,Hillery:1984oq,Balazs:1984cr,Lee:1995tg,Schroek:1996fv,Ozorio:1998aa,Schleich:2001hc,QMPS:2005mi,Weinbub:2018aa,Rundle:2021wd}  where observables become $c$-number functions (the \textit{symbols} of the operators). This is the most convenient construct for visualizing quantum states and processes for continuous variables (CV). 

An alternative approach well suited to compare the classical and quantum evolutions of a given system is to decompose the wave function (or density matrix) into its infinite set of statistical moments. These moments contain the same physical information as the wave function, but they have the advantage of being observable. Their evolution equations give rise to Hamiltonian equations with quantum corrections coming from momentum variables~\cite{Andrews:1985aa,Ballentine:1998aa,Bojowald:2006aa,Brizuela:2014aa,Ahmadzadegan:2016aa,Andrews:2021aa,Andrews:2022aa}. It seems thus natural to look at the monomials
\begin{equation}
\hat{T}_{Kq}= \had^{K+q} \ha^{K-q}
\label{eq:opten}
\end{equation} 
with $K=0, 1/2, 1, \ldots$ and $q= - K, \ldots, + K$. Here, we restrict ourselves to a single mode, with bosonic creation and annihilation operators $\ha$ and $\had$, respectively. The extension to multiple bosonic modes is direct.

We explore here the properties of this basis and check its explicit   invariance under symplectic transformations (i.e., linear canonical transformations), which is not apparent at first sight~\cite{Englert:1989aa}. Some work along these lines can be found in Ref.~\cite{Ivan:2012aa}, but restricted to the symmetric (or Weyl) ordering. Here, we examine the inverse of these monomials for arbitrary operator orderings, so they can   be used to directly expand any quantum operator, as it would be required for proper tensorial sets for CV. These operators can then be added to the quantum optician's toolbox and used by anyone working in CV.

When an arbitrary pure or mixed density matrix is expanded in the basis \eqref{eq:opten}, its expansion coefficients are the moments, dubbed as state multipoles, which convey complete information. These moments couple parts of the density matrix corresponding to different particle numbers, unlike $2K$-particle density matrix expansions in quantum statistical mechanics, although some of the $q=0$ moments appear in both contexts~\cite{Kummer:1970aa,Ter-Haar:1961aa,Bogolubov:1970aa}. For CV, moments have been considered for studying quantumness~\cite{Shchukin:2005aa,Shchukin:2005ab}. Here, we inspect how the multipoles characterize the state. Drawing inspiration from SU(2), we compare states that hide their information in the large-$K$ coefficients to those whose information is mostly contained in the smallest-$K$ multipoles. The result is an intriguing counterplay between the extremal states in the other representations, including Fock states, coherent states, and states with maximal off-diagonal coefficients in the Fock basis.  

There are many avenues to explore with the monomials representation. After a brief review of the basic concepts required in Sec.~\ref{sec:back}, we examine the properties of the basis \eqref{eq:opten} and its inverse in Sec.~\ref{sec:invmon}. The corresponding multipoles appear as the expansion coefficients of the density matrix in that basis. The covariance under symplectic transformations tells  us how the different parts of a state are interconverted through standard operations.  Note that we are considering only normally ordered polynomials, but everything can be extended for antinormally and symmetrically ordered monomials. In Sec.~\ref{sec:Ext} we introduce the concept of cumulative multipole distribution and its inverse and find the extremal states for those quantities and determine in this way which states are the most and least quantum. The direct connections between our multipoles, tomography, quantization, and quasiprobability distributions are elucidated in Sec.~\ref{sec:applications}. Our conclusions are finally summarized in Sec.~\ref{sec:conc}.

\section{Background}
\label{sec:back}

We provide here a self-contained background that is familiar to quantum opticians. The reader can find more details in the previously quoted literature~\cite{Tatarskii:1983uq,Hillery:1984oq,Balazs:1984cr,Lee:1995tg,Schroek:1996fv,Ozorio:1998aa,Schleich:2001hc,QMPS:2005mi,Weinbub:2018aa,Rundle:2021wd}. 
A single bosonic mode has creation and annihilation operators satisfying the commutation relations
\begin{equation}
 [\ha, \had  ]= \hat{\openone} .
\end{equation} 
These can be used to define the Fock states as excitations
\begin{equation}
\ket{n} = \frac{\had^n}{\sqrt{n!}}\vac
\end{equation} 
of the vacuum $\vac$ annihilated as $\ha\vac=0$, as well as the canonical coherent states
\begin{equation}
\ket{\alpha}=\eu^{-\tfrac{|\alpha|^2}{2}}\sum_{n=0}^\infty\frac{\alpha^n}{\sqrt{n!}}\ket{n}.
\end{equation}
These can both be used to resolve the identity:
\begin{equation}
\hat{\openone}=\sum_{n=0}^\infty \ket{n}\bra{n}=\frac{1}{\pi}\int d^2\alpha\ket{\alpha}\bra{\alpha}.
\end{equation}

The coherent states can also be defined as displaced versions of the vacuum state $\ket{\alpha}=\hat{D}(\alpha)\vac$ via the displacement operators that take numerous useful forms
\begin{equation}
\hat{D}( \alpha )= e^{\alpha \had - \alpha^{\ast} \ha} = 
e^{-\tfrac{|\alpha|^2}{2}} \, e^{\alpha \had} e^{-\alpha^{\ast} \ha} =
e^{\tfrac{|\alpha|^2}{2}} \,  e^{-\alpha^{\ast} \ha} e^{\alpha \had} \, .
\end{equation}
These obey the composition law
\begin{equation}
\hat{D} (\alpha ) \hat{D} ( \beta ) = e^{i \im (\alpha \beta^{\ast})} \hat{D} ( \alpha + \beta ) 
\end{equation}
and the trace-orthogonality condition
\begin{equation}
\Tr [ \hat{D} ( \alpha ) \hat{D} (- \beta ) ]=\pi \delta^2(\alpha-\beta) \, .
\end{equation} 
Their matrix elements in the coherent-state basis can be found from the composition law and in the Fock-state basis are given by~\cite{Perelomov:1986aa}  
\begin{equation}
\bra{m}\hat{D} (\alpha ) \ket{n} = 
\begin{cases}
\displaystyle \sqrt{\frac{n!}{m!}} e^{-\tfrac{|\alpha|^2}{2}} \alpha^{m-n} \, L_n^{(m-n)}(|\alpha|^2),&m\leq n , \\
& \\
\displaystyle \sqrt{\frac{m!}{n!}} e^{-\tfrac{|\alpha|^2}{2}}(-\alpha^{\ast})^{n-m}L_m^{(n-m)}(|\alpha|^2),&n\leq m ,
\end{cases}
\end{equation}
where $L_{n}^{(\alpha)} ( \cdot )$ denotes the generalized Laguerre polynomial~\cite{NIST:DLMF}.

Given any operator $\hat{F}$,  it can be expressed in the Fock basis as
\begin{equation}
\hat{F}=\sum_{m,n} F_{m,n}\ket{m}\bra{n}, \qquad F_{m,n}=\bra{m}\hat{F}\ket{n}
\end{equation} 
and in the coherent-state basis as
\begin{equation}
\hat{F}=\frac{1}{\pi^2} \int d^2\alpha d^2\beta  F (\alpha,\beta ) \ket{\alpha}\bra{\beta},
\qquad F\left(\alpha,\beta\right)=\bra{\alpha}\hat{F}\ket{\beta}.
\end{equation} 
However, it is always possible to express this coherent-state representation in a  diagonal form. For the particular case of the density operator $\hat{\varrho}$ this yields the Glauber-Sudarshan $P$-function~\cite{Glauber:1963aa,Sudarshan:1963aa}
\begin{equation}
\hat{\varrho} = \int d^2\alpha \, P (\alpha )\ket{\alpha}\bra{\alpha} \, ,
\end{equation}
with~\cite{Mehta:1967aa}
\begin{equation}
P (\alpha ) = \frac{e^{|\alpha|^2}}{\pi^2}\int d^2\beta \bra{-\beta}\hat{\varrho}\ket{\beta}
e^{|\beta|^2 + 2 i \im (\alpha\beta^{\ast})} .
\end{equation}
The same holds true for any operator $\hat{F}$ for which $\bra{-\beta}\hat{F}\ket{\beta} e^{|\beta|^2}$ is square-integrable.

One identity that often shows up in this realm is an expression for the vacuum in terms of normally ordered polynomials~\cite{Cahill:1969aa}:
\begin{equation}
\vac\bra{\mathrm{vac}} = :e^{-\had\ha}: \, .
\end{equation}
This allows us to express any unit-rank operator from the Fock basis as
\begin{equation}
\ket{m}\bra{n}= \frac{1}{\sqrt{m!n!}} :\had^m\eu^{-\had\ha}\ha^n: =\frac{1}{\sqrt{m!n!}}\sum_{k=0}^\infty \frac{(-1)^k}{k!} \had^{m+k}\ha^{n+k}.
\label{eq:normal ordering expression for Fock}
\end{equation} 
Since any operator can be written as a linear combination of operators $|m\rangle\langle n|$, this directly guarantees that a normally ordered expression will always exist for any operator.

\section{State multipoles}
\label{sec:invmon}

As heralded in the Introduction, the monomials \eqref{eq:opten} are the components of finite-dimensional tensor operators with respect to the symplectic group Sp(2, $\mathbb{R}$). Their transformation properties are examined in the Appendix~\ref{sec:apen}.  For completeness, we have to seek operators $\hat{\mathfrak{T}}_{Kq}$ satisfying the proper orthonormality conditions to be inverses of the monomials:
\begin{equation}
\Tr ( \hat{\mathfrak{T}}_{Kq} \hat{T}_{K^\prime q^\prime} ) = 
\delta_{K K^\prime}\delta_{q q^\prime} \, .
\end{equation}
Using the trace-orthogonality conditions of the displacement operators, we can rewrite this condition as
\begin{align}
 \Tr (\hat{\mathfrak{T}}_{Kq}\hat{T}_{K^\prime q^\prime} ) & = \frac{1}{\pi} \int d^2\beta \,
 \Tr[\hat{D} (\beta )\hat{T}_{K^\prime q^\prime}] \Tr [ \hat{D} (-\beta ) \hat{\mathfrak{T}}_{K q}] \nonumber \\
 &= \frac{1}{\pi^{2}} \int d^2\alpha d^2\beta \, e^{\tfrac{|\beta|^2}{2}}  e^{\beta\alpha^{\ast}-\beta^{\ast}\alpha} \alpha^{\ast K^\prime +q^\prime} \alpha^{K^\prime -q^\prime} 
 \Tr [\hat{D} (-\beta )\hat{\mathfrak{T}}_{K q} ] \, .
\end{align}
Now, by inspection, we attain orthonormality when 
\begin{equation}
e^{\tfrac{|\beta|^2}{2}} \Tr [\hat{D} (-\beta ) \hat{\mathfrak{T}}_{K q} ]= (-1 )^{2K} 
\frac{\beta^{K+q} (-\beta^{\ast})^{K-q}}{(K+q)! (K-q)!} \, ,
\label{eq:Tr D inv}
\end{equation}
because then the derivatives of the delta function 
\begin{equation} 
\frac{\partial^{2K}}{\partial \beta^{K+q}\partial (-\beta^*)^{K-q}} \delta^2(\beta)= \frac{1}{\pi^{2}} \int d^2\alpha \;  e^{\beta\alpha^{\ast}-\beta^{\ast}\alpha}\alpha^{*K+q}\alpha^{K-q}
\end{equation} 
will match perfectly when performing integration by parts. In consequence, we have
\begin{equation}
\hat{\mathfrak{T}}_{K q} =\frac{\left(-1\right)^{K+q}}{\left(K+q\right)!\left(K-q\right)!} \frac{1}{\pi} \int d^{2}\beta \; e^{-\tfrac{|\beta|^2}{2}}\hat{D} (\beta ) \; \beta^{K+q}\beta^{\ast K-q}.
\label{eq:inverse operators integrals}
\end{equation}
Interestingly, they appear as moments of the operators introduced in the pioneering work by Agarwal and Wolf~\cite{Agarwal:1970aa}. This inversion process can be repeated with other ordered polynomials and we find the inverse operators to again appear as moments of the other operators introduced therein. In Appendix~\ref{sec:symord} we sketch the procedure for the case of symmetric order, where our technique is shown to be useful for finding the inverse operators for arbitrary operator ordering, whenever such exist. Once they are known, it is easy to expand any operator, such as a density matrix $\hat{\varrho}$, through
\begin{equation}
\hat{\varrho}= \sum_{Kq} \langle \hat{\mathfrak{T}}_{Kq} \rangle \; \hat{T}_{Kq} \, ,
\end{equation}
where $\langle \hat{\mathfrak{T}}_{Kq} \rangle  = \Tr (\hat{\varrho} \hat{\mathfrak{T}}_{Kq})$, following the standard notation for SU(2)~\cite{Blum:1981aa}, will be called the state multipoles. They correspond to moments of the basic variables, properly arranged.

Conversely, we can expand operators in the basis of the inverse operators,  
\begin{equation}
\hat{\varrho}= \sum_{Kq} \langle \hat{T}_{Kq} \rangle \; \hat{\mathfrak{T}}_{Kq} \, ,
\label{eq:rho from inverse}
\end{equation} 
with $ \langle \hat{T}_{Kq} \rangle = \Tr (\hat{\varrho} {T}_{Kq}) $ now being the inverse multipoles.

Since inverse operators inherit the Hermitian conjugation properties of the monomials,
\begin{equation}
\hat{T}_{Kq}^\dagger= \hat{T}_{K \, - q} , \qquad \qquad
\hat{\mathfrak{T}}_{Kq}^\dagger = \hat{\mathfrak{T}}_{K \, -q},
\end{equation} 
the multipoles and inverse multipoles simply transform as $q\leftrightarrow -q$ under complex conjugation. 

The purity of a state has a simple expression in terms of the multipoles
\begin{equation}
\Tr(\hat{\varrho}^2) = \sum_{Kq} \langle \hat{\mathfrak{T}}_{Kq} \rangle \langle \hat{T}_{Kq} \rangle  \, .
\end{equation} 
It is more challenging to express the trace of a state in terms of the multipoles because the operators $\hat{T}_{Kq}$ are not trace-class; however, by formally writing $\Tr[\hat{D}(\beta)]=\pi\delta^2(\beta)\exp(-|\beta|^2/2)$, we can compute
\begin{equation}
\Tr(\hat{\mathfrak{T}}_{Kq}) = \delta_{K0}\delta_{q0}
\end{equation} 
such that normalization dictates that the inverse multipoles satisfy $1 = \Tr(\hat{\varrho})= \langle \hat{T}_{00} \rangle$.

Extension of all of these results to the multimode case merely requires concatenation of multipole copies of our results, since operators from different bosonic modes commute with each other. Considering the extended monomial operators 
\begin{equation}
    \hat{T}_{K_1\cdots K_N q_1\cdots q_N} =\ha_1^{\dagger K_1+q_1}\cdots \ha_N^{\dagger K_N+q_N} \ha_1^{K_1-q_1}\cdots  \ha_N^{K_N-q_N},
\end{equation} the extended inverse operators are obtained from our inverse operators as
\begin{equation}
    \hat{\mathfrak{T}}_{K_1\cdots K_N q_1\cdots q_N}=\hat{\mathfrak{T}}_{K_1q_1}\cdots \hat{\mathfrak{T}}_{K_Nq_N}.
\end{equation} 
The extended orthonormality conditions are simply
\begin{equation}
\Tr ( \hat{\mathfrak{T}}_{K_1\cdots K_Nq_1\cdots q_N} \hat{T}_{K_1^\prime\cdots K_N^\prime q_1^\prime\cdots q_N^\prime} ) = 
\delta_{K_1 K_1^\prime}\delta_{q_1 q_1^\prime}\cdots \delta_{K_N K_N^\prime}\delta_{q_N q_N^\prime},
\end{equation}
which immediately yields results such as the purity of a multimode state being
\begin{equation}
\Tr(\hat{\varrho}^2) = \sum_{K_1\cdots K_Nq_1\cdots q_N} \langle \hat{\mathfrak{T}}_{K_1\cdots K_Nq_1\cdots q_N} \rangle \langle \hat{T}_{K_1\cdots K_Nq_1\cdots q_N} \rangle  \, .
\end{equation} 
All such generalizations to the multimode case are similarly straightforward so we proceed by only focusing on the essential computations for a single mode.

In principle, the complete characterization of a CV state requires the knowledge of infinite multipoles. For a Gaussian state, only moments up until $K=1$ are needed , as these encode all of the means and covariances of position and momentum operators, which are the only relevant degrees of freedom in a Gaussian state. This suggests that either the inverse multipoles $\langle \hat{{T}}_{Kq} \rangle $ for larger values of $K$ or the multipoles $\langle \hat{\mathfrak{T}}_{Kq} \rangle $ characterize the non-Gaussianity of a state.

\begin{figure}
    \centering
    \includegraphics[width=0.95\columnwidth]{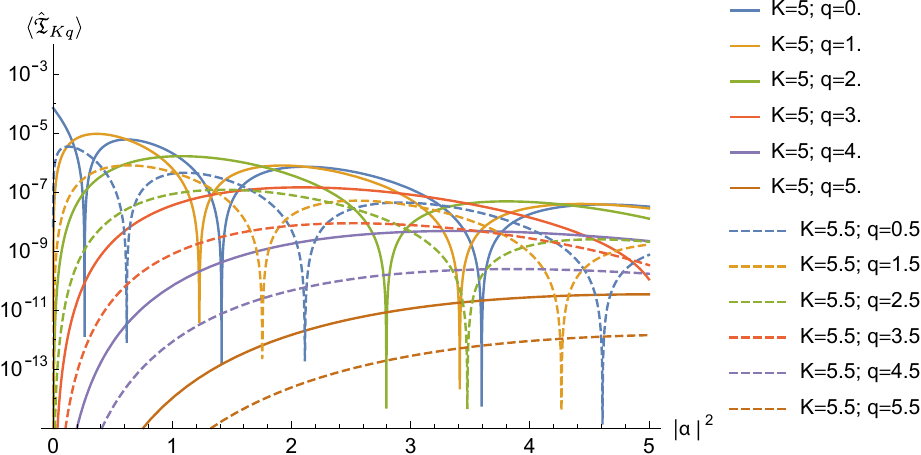}
    \caption{Multipole moments $\langle \alpha|\hat{\mathfrak{T}}_{Kq}|\alpha\rangle$ of order $K=5$ and $K=11/2$ for coherent states with average energy $|\alpha|^2$. Note that $|\langle \alpha|\hat{\mathfrak{T}}_{Kq}|\alpha\rangle|= |\langle \alpha|\hat{\mathfrak{T}}_{K -q}|\alpha\rangle|$ and that these diminish with increasing $|q|$, $|\alpha|^2$, and $K$. The moments are depicted in a logarithmic scale; the occasional sharp dips occur at the zeroes of the Lauerre polynomials.}
    \label{fig:coh multipoles K5}
\end{figure}

In consequence, we have to calculate the multipoles of arbitrary states. Before that, we consider the simplest cases of coherent and Fock states, for which the calculations are straightforward.  Starting with coherent states, using \eqref{eq:inverse operators integrals} and recalling the Rodrigues formula for the generalized Laguerre polynomials~\cite{NIST:DLMF}, we get
\begin{equation}
\begin{aligned}
    \langle \alpha | \hat{\mathfrak{T}}_{Kq} |\alpha \rangle 
&=\frac{\left(-1\right)^{K+q}}{\left(K+q\right)!\left(K-q\right)!} \frac{1}{\pi} \frac{\partial^{2K}}{\partial\alpha^{K-q}\partial(-\alpha^*)^{K+q}}\int d^{2}\beta \; e^{-|\beta|^2+\alpha\beta^*-\alpha^*\beta}
\\&= \frac{(-1)^{K+q}}{(K-q)!}\frac{e^{-|\alpha|^2}}{\alpha^{\ast 2q}}
L_{K+q}^{(-2q)}(|\alpha|^2) \, . 
\end{aligned}
\end{equation} 
The magnitudes of these multipole moments versus $|\alpha|^2$ for various values of $K$ and $q$ are plotted in Fig.~\ref{fig:coh multipoles K5}. As we can appreciate, they decrease rapidly with $K$ and large $|\alpha|$, occasionally vanishing at the values of $|\alpha|$ for which a particular Laguerre polynomial vanishes. The overall structure follows the Laguerre polynomials diminished by the exponentially decaying function $\exp(-|\alpha^2)$ and the factorial factor $1/(K-q)!$. We see that, even for states with a large amount of energy, the majority of the information may be contained in the lower-order moments.

As for Fock states, we use the matrix elements of the displacement operator $\bra{n}\hat{D}(\beta)\ket{n}=\exp(-|\beta|^2/2)L_n(|\beta|^2)$. Since these only depend on $|\beta|$ and not its phase, the $q=0$ terms all vanish, leaving us with
\begin{equation}
\langle n | \hat{\mathfrak{T}}_{Kq} |n\rangle=\delta_{q0}\frac{(-1)^K}{K!^2}\int_0^\infty r dr \,  2\eu^{-r^2}r^{2K}L_n(r^2) = \delta_{q0}\frac{(-1)^{K+n}}{n!(K-n)!}.
\end{equation}
The inverse multipoles are trivial in both cases, with
\begin{equation}
\langle \alpha | \hat{T}_{Kq} |\alpha \rangle  = \alpha^{\ast K+q}\alpha^{K-q} \,, 
\qquad \qquad
\langle n | \hat{T}_{Kq} |n\rangle =\delta_{q0}K!\binom{n}{K} \, .
\end{equation} 
Note that the multipoles that vanish for Fock states have $n>K$ and the inverse multipoles that vanish for Fock states have $K>n$.

\begin{figure}[t]
    \centering
    \includegraphics[width=0.5\columnwidth]{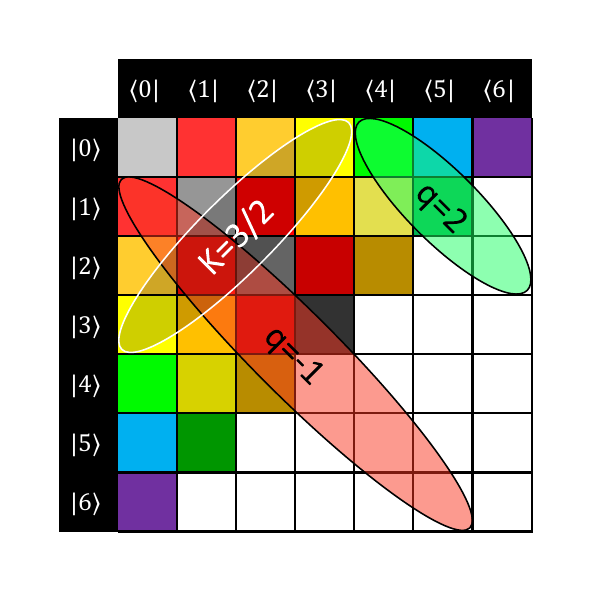}
    \caption{Parts of the state in the Fock state basis coupled to by a particular inverse operator $\hat{\mathfrak{T}}_{Kq}$. Each value of $q$ labels the off-diagonal stripe (grouped with a particular colour) of the matrix that affects the value of $\langle \hat{\mathfrak{T}}_{Kq}\rangle$. For example, when computing $\langle \hat{\mathfrak{T}}_{K2}\rangle$, one uses only components of the density matrix in the $q=2$ oval of the state's coefficients when represented in the Fock basis. Each value of $K$ labels the maximal antidiagonal row that contributes to the value of $\langle \hat{\mathfrak{T}}_{Kq}\rangle$. This antidiagonal row is characterized by the row and column number summing to $2K$ (the colours get darker as the antidiagonal increases from the top left). For example, when computing $\langle \hat{\mathfrak{T}}_{\tfrac{3}{2}2}\rangle$ one looks at the intersection of the $K=3/2$ and $q=2$ ovals and finds no overlap, such that there is no contribution from any state, while computing $\langle \hat{\mathfrak{T}}_{\tfrac{3}{2}-1}\rangle$ requires the coefficients of $|2\rangle\langle 1|$ and $|1\rangle\langle 0|$ when the density matrix is expanded in the Fock basis, as per the $q=-1$ oval telling us which components might contribute and the $K=3/2$ oval telling us the limit beyond which no elements contribute.}
    \label{fig:parts_of_matrix}
\end{figure}

For arbitrary states, we note that, since any state can be expressed in terms of its $P$-function, we can write
\begin{equation}
\langle \hat{\mathfrak{T}}_{Kq} \rangle =\int d^2\alpha  \; P(\alpha) \langle \alpha | \hat{\mathfrak{T}}_{Kq} |\alpha \rangle = \int d^2\alpha  \; P(\alpha) \, \frac{(-1)^{K+q}}{(K-q)!}\frac{e^{-|\alpha|^2}}{\alpha^{\ast 2q}}L_{K+q}^{(-2q)}(|\alpha|^2).
\end{equation} 
To get more of a handle on these multipoles, expecially when $P$ is not a well-behaved function, it is more convenient to have an expression in terms of the matrix elements $\varrho_{mn}=\bra{m}\hat{\varrho}\ket{n}$. This can be provided by expressing $P(\alpha)$ in terms of matrix elements of the state in the Fock basis and derivatives of delta functions. More directly, we can compute ($m\leq n$)
\begin{align}
\langle n | \hat{\mathfrak{T}}_{Kq} | m \rangle &=\frac{\left(-1\right)^{K+q}}{\left(K+q\right)!\left(K-q\right)!} 
\frac{1}{\pi} \int d^2\beta e^{-|\beta|^2}\sqrt{\frac{n!}{m!}}\beta^{m-n}L_n^{(m-n)}(|\beta|^2)\beta^{K+q}\beta^{\ast K-q} \nonumber \\
&=\delta_{n-m,2q}\left(-1\right)^{K+q+n}\sqrt{\frac{n!}{(n-2q)!}}\binom{K+q}{n}\frac{1}{(K+q)!}.
\label{eq:multipoles for Fock rank one}
\end{align}
These give the matrix elements of the inverse operators $\hat{\mathfrak{T}}_{Kq}$ in the Fock basis and show that $\hat{\mathfrak{T}}_{Kq}$ can only have nonnull eigenstates when $q=0$. Putting these together for an arbitrary state, we find
\begin{equation}
\langle \hat{\mathfrak{T}}_{Kq}  \rangle = \begin{cases} 
\displaystyle \sum_{n\geq m}\varrho_{nm}\delta_{n-m,2q}\left(-1\right)^{K+q+n}\sqrt{\frac{n!}{(n-2q)!}}\binom{K+q}{n}\frac{1}{(K+q)!},& q\geq 0 \, , \\
& \\
\displaystyle \sum_{m\geq n}\varrho_{nm}^*\delta_{n-m,-2q}\left(-1\right)^{K-q+n}\sqrt{\frac{n!}{(n+2q)!}}\binom{K-q}{n}\frac{1}{(K-q)!},& q\leq 0 \, .
\end{cases}
\label{eq:multipole moments in terms of fock basis}
\end{equation}
In this way, we get a simple expression for the inverse monomials in the Fock basis:
\begin{equation}
\hat{\mathfrak{T}}_{Kq} =\begin{cases}
\displaystyle 
\sum_{n=2q}^{K+q} \frac{(-1)^{K+q+n}}{\sqrt{n!(n-2q)!}(K+q-n)!} \, \ket{n-2q}\bra{n} ,& 
q \geq 0 \, , \\
& \\
\displaystyle
\sum_{n=-2q}^{K-q}\frac{(-1)^{K-q+n}}{\sqrt{n!(n+2q)!}(K-q-n)!} \, \ket{n}\bra{n+2q} ,& 
q\leq 0 \, ,
\end{cases}
\label{eq:inverse operators Fock basis}
\end{equation} 
whose orthonormality with the operators $\hat{T}_{Kq}$ can be directly verified. This expression equally serves to furnish a representation of the moments of the displacement operator in the Fock basis. 

To understand this result, we plot in Fig.~\ref{fig:parts_of_matrix}  a representation of the nonzero parts of different operators $\hat{\mathfrak{T}}_{Kq}$ in the Fock basis, which equivalently represents which elements of a density matrix $\varrho_{mn}$ contribute to a given multipole $\langle \hat{\mathfrak{T}}_{Kq} \rangle$. The contributing elements are all on the $2q$th diagonal, ranging over the first $2K+1$ antidiagonals. The inverse multipoles $\langle \hat{T}_{Kq} \rangle$ depend on the $-q$th diagonal and all of the antidiagonals starting from the $2K$th antidiagonal. These are in contrast with expansions in quantum statistical mechanics that only retain information about the parts of a state with a fixed number of particles (here $q=0$). This picture makes clear a number of properties that will become useful for our purposes.

To conclude, it is common to find operators of a generic form $f(\ha,\had)$. Quite often, it is necessary to find their normally ordered form $\normord{f(\ha,\had)}$,  where $: \, :$ denotes normal ordering.  Such is necessary, for example, in photodetection theory~\cite{Sperling:2012aa}. Although algebraic techniques are available~\cite{Louisell:1973aa}, the multipolar expansion that we have developed makes this computation quite tractable. We first compute
\begin{equation}
\Tr[\hat{D}(\beta)\,\normord{f(\ha,\had)}]= e^{\frac{|\beta|^2}{2}} \,  
\Tr[\normord{e^{\beta\had} \, f(\ha,\had) \, e^{-\beta^{\ast} \ha}}] = \frac{e^{\frac{|\beta|^2}{2}}}{\pi} \int d^2\alpha \, f(\alpha,\alpha^{\ast}) \, e^{\beta \alpha^{\ast}-\beta^{\ast} \alpha} \, .
\end{equation} 
The integral is nothing but the Fourier transform (note $\beta\alpha^*-\beta^*\alpha$ is purely imaginary) of the function $f(\alpha,\alpha^{\ast})$ with respect to both of its arguments. If we call $F(\beta, \beta^{\ast})$ this transform, the  multipole moments of  $:f(\ha,\had):$, denoted by $F_{Kq}$, become
\begin{equation}
F_{Kq}= \frac{(-1)^{K+q}}{\pi (K+q)! (K-q)!} \int d^2\beta \, F(\beta,\beta^{\ast}) \, \beta^{K+q}\beta^{\ast K-q} \, .
\end{equation}
In other words, the moments of the Fourier transform of $f(\alpha,\alpha^{\ast})$ give the expansion coefficients of $:f(\ha,\had):$ in the $\hat{{T}}_{Kq}$ basis.

\section{Extremal states}
\label{sec:Ext}

\subsection{Cumulative multipolar distribution}

We turn now our attention to cumulative multipole distribution, which in the context of SU(2) is a good quantifier of quantumness, and inspect whether the cumulative multipoles are good quantifiers of quantumness for CV as well. That is, we form the cumulative multipole distributions
\begin{equation}
\mathfrak{A}_M (\hat{\varrho}) = \sum_{K=0}^M \mathfrak{T}_{K}^{2}(\hat{\varrho})
\label{eq:cum multipole defn}
\end{equation} 
with  $M=0, 1/2, 1, \ldots$ and where
\begin{equation}
\mathfrak{T}_{K}^{2}(\hat{\varrho}) = \sum_{q=-K}^{K}  |\Tr (\hat{\mathfrak{T}}_{Kq} \hat{\varrho}) |^{2} 
\end{equation}
is the Euclidean norm of the $K$th multipole. The quantities $\mathfrak{A}_M (\hat{\varrho})$ can be be used to furnish a generalized uncertainty principle for CV~\cite{Ivan:2012aa} and they are a good indicator of quantumness~\cite{Goldberg:2020aa,Goldberg:2022aa}. For spin variables, it has been shown that  $\mathfrak{A}_{M} (\hat{\varrho})$  are maximized to all orders $M$ by SU(2)-coherent states, which are the least quantum states in this context, and vanish for the most quantum states, which are called the Kings of Quantumness, the furthest in some sense from coherent states~\cite{Hoz:2013om,Hoz:2014kq,Bjork:2015aa}. 

What states maximize and minimize these cumulative variables for CV? Do these extremal states behave as for SU(2), where they are respectively the least and most quantum states? Can we use the cumulative multipole moments as a proxy for inspecting the quantumness of a state, as can be done in SU(2)? Let us begin by examining a few of the lowest orders.

$\bm{M=0}$: For an arbitrary state, we can write $\mathfrak{A}_{0}$ in terms of the Fock-state coefficients as
\begin{equation}
\mathfrak{A}_{0} = \left |\sum_n(-1)^{n} \varrho_{nn} \binom{0}{n}\right|^2 =  
|\varrho_{00} |^{2} .
\end{equation}
This is uniquely maximized by the vacuum state $\vac$, with $\varrho_{00}=1$, which is a minimal-energy coherent state and can be considered the least quantum state in this context. The quantity $\mathfrak{A}_{0}$, on the other hand, is minimized by any state with $\varrho_{00}=0$, which causes $\mathfrak{A}_0$ to vanish. This is easily attained by Fock states $\ket{n}$ with $n>0$. In this sense, all Fock states that are not the vacuum are the most quantum. States becomes more quantum as they gain more energy and their vacuum component $\varrho_{00}$ diminishes in magnitude.

$\bm{M=1/2}$: For $K=1/2$, we can readily compute
\begin{equation}
\mathfrak{T}_{1/2}= |\varrho_{01}|^2 + |\varrho_{10}|^2 = 2 |\varrho_{01}|^2.
\end{equation} 
This is minimized by any state with no coherences in the Fock basis (such as, e.g., number states). On the other hand, it is maximized by states with maximal coherence in the smallest-energy section of the Fock basis: $\ket{\psi_+} = \tfrac{1}{\sqrt{2}} (\ket{0} + e^{i \varphi}\ket{1})$, with $\varphi\in\mathbb{R}$. Together, $\mathfrak{A}_{1/2}$ is minimized by any state with $\varrho_{00}=0$, because that forces $\varrho_{01}$ to vanish by positivity of the density matrix, and it is still uniquely maximized by the vacuum state, again because of the positivity constraint $|\varrho_{01}|\leq\sqrt{\varrho_{00}(1-\varrho_{00})}$. 

$\bm{M=1}$: Now, we find
\begin{equation}
\mathfrak{T}_{1} = |\varrho_{00}-\varrho_{11} |^2+ \tfrac{1}{2} |\varrho_{02}|^2 + \tfrac{1}{2} |\varrho_{20} |^2 = (\varrho_{00}-\varrho_{11})^2 + |\varrho_{02} |^2.
\end{equation}
This is minimized by all states with $\varrho_{00}=\varrho_{11}=0$, again including Fock states but now with more than one excitation, but it is also \textit{minimized} by the state $\ket{\psi_+}$ that \textit{maximized} $\mathfrak{A}_{1/2}$. It is again maximized by the vacuum state with $\varrho_{00}=1$, but it is also maximized by the single-photon state with $\varrho_{11}=1$. The cumulative distribution is again the more sensible quantity: $\mathfrak{A}_1$ is minimized by states with vanishing components in the zero- and single-excitation subspaces, of which the Fock state $\ket{2}$ has the lowest energy, and is uniquely maximized by the vacuum (coherent) state. 

$\bm{M=3/2}$: We find
\begin{equation}
\mathfrak{T}_{3/2}= \tfrac{2}{3!} |\varrho_{30} |^2 + 2 \left |\varrho_{10}- \tfrac{1}{\sqrt{2}} \varrho_{21}\right|^2.
\end{equation} 
As usual this is minimized by any Fock state and by any state with no probability in photon-number sectors up until $n=3$, while it is maximized by pure states of the form $\ket{\psi} = e^{i \varphi} \tfrac{1}{\sqrt{3}}\ket{0}+\tfrac{1}{\sqrt{2}}\ket{1}-e^{-i \varphi}\tfrac{1}{\sqrt{6}}\ket{2}$. The cumulative $\mathfrak{A}_{3/2}$ is again uniquely maximized by the vacuum state and minimized by any Fock state and by any state with no probability in photon-number sectors up until $n=3$.

$\bm{M>3/2}$: The consistent conclusion is that different Euclidean norms of the multipoles for different orders $K$ can be maximized by different states, but that the cumulative distribution is always maximized by the vacuum state. All of the orders of multipoles and their cumulative distribution vanish for sufficiently large Fock states, cementing Fock states as maximally quantum according to this condition. We as of yet have only a circuitous proof that  $\mathfrak{A}_{M} (\hat{\varrho})$  is uniquely maximized by $\vac$ for arbitrarily large $M$: in Appendix \ref{app:vac max}, we provide joint analytical and numerical arguments that this pattern continues for all $M$, such that the vacuum state may be considered minimally quantum according to this condition.

We can compute this maximal cumulative multipole moment, that of the vacuum, at any order: 
\begin{equation}
\mathfrak{A}_M(\vac)=\sum_{K=0}^M\frac{1}{K!^2}=I_0(2)-\, _1\tilde{F}_2(1;\lfloor M \rfloor+2,\lfloor M \rfloor+2;1),
\end{equation} 
with a Bessel function~\cite{NIST:DLMF} and a regularized hypergeometric function~\cite{Weisstein:2023aa}. This approaches $I_0(2)\approx 2.27959$ in the limit of large $M$. Moreover, by computing $\mathfrak{A}_\infty(\ket{n})=I_0(2)/n!^2$, we realize why only $\ket{0}$ and $\ket{1}$ behave so similarly in the large-$M$ limit.

Finally, note that the cumulative multipole operators also take the intriguing form
\begin{align}
\hat{\mathfrak{A}}_M &= \displaystyle \frac{1}{\pi^{2}} \int d^2\alpha d^2\beta \, e^{-\tfrac{|\alpha|^2+|\beta|^2}{2}}\hat{D} (-\alpha)\otimes\hat{D}(\beta) \sum_{K}^M \frac{\left(\alpha \beta^{\ast} -\alpha^{\ast} \beta \right)^{2K}}{(2K)!^2}  P_{2K} \left(\frac{\alpha \beta^{\ast} +\alpha^{\ast}\beta}{\alpha^{\ast} \beta- \alpha\beta^{\ast}}\right) \nonumber \\
& \\
\hat{\mathfrak{A}}_\infty &= \displaystyle \frac{1}{\pi^{2}} \int d^2\alpha d^2\beta  \, e^{-\tfrac{|\alpha|^2+|\beta|^2}{2}} \hat{D}(-\alpha)\otimes\hat{D}(\beta) \left| I_0(2\sqrt{\alpha \beta^{\ast}})\right|^2 \nonumber ,
\end{align} 
where $P_n(\alpha)= \exp{- |\alpha|^2/2}\alpha^n/\sqrt{n!}$ is the Poissonian amplitude. 

We thus conclude that the multipole moments defined here are a proxy for the ``vacuum-stateness'' or ``Fock-stateness'' of a quantum state. Like their SU(2) counterparts, these multipoles can be used to quantify  quantumness, with the most quantum states having the lowest cumulative multipole moments and vice versa. Since in many known contexts the vacuum states and Fock states define the limits of the least and most quantum states~\cite{Goldberg:2020aa}, this implies that measuring the lowest-order multipoles could already be well defined for inspecting the quantumness of a quantum state.

\subsection{Inverse multipole distribution}

An important question arises: how does one measure a state's multipole moments? Homodyne detection provides one immediate answer. By interfering a given state $\hat{\varrho}$ with a coherent state $\ket{\alpha}$ on a balanced beamsplitter and measuring the difference of the photocurrents of detectors placed at both output ports, one collects a signal proportional to $ x(\theta)=\expct{\ha e^{- i \theta}+\had e^{i \theta}}$, where $\theta$ can be varied by changing the phase $\arg \alpha$ of the reference beam. Collecting statistics of the quadrature $x(\theta)$ up to its $K$th-order moments for a variety of phases $\theta$ allows one to read off the moments $\langle \hat{T}_{Kq} \rangle = \expct{\had^{K+q}\ha^{K-q}}$. This provokes the question: what states maximize and minimize the cumulative multipole moments in the inverse basis?

We start by defining, in analogy to Eq.~\eqref{eq:cum multipole defn}, the cumulative distribution
\begin{equation}
{A}_M (\hat{\varrho})=\sum_{K}^M\sum_{q=-K}^K \left| \langle \hat{T}_{Kq} \rangle \right|^2 \, .
\end{equation} 
This quantity directly depends on the energy of the state, vanishing if an only if the state is the vacuum.  As for the maximization, it is clear that coherent states with more energy cause the cumulative sum ${A}_M$ to increase, so we must fix the average energy $\bar{n}=\expct{\had\ha}$ when comparing which states maximize and minimize the sum.

Maximizing $A_M$ for a fixed average energy is straightforward because each inverse multipole satisfies
\begin{equation}
| \langle \hat{T}_{Kq} \rangle |^{2} \leq \expct{\had^{K+q}\ha^{K+q}} \expct{\had^{K-q}\ha^{K-q}}.
\end{equation} 
The inequality is saturated if and only if $\ha^{K+q}\ket{\psi}\propto\ha^{K-q}\ket{\psi}$; that is, $\ha^{2q}\ket{\psi}\propto\ket{\psi}$, which, for $q\neq 0$, requires coherent states or superpositions of coherent states with particular phase relationships akin to higher-order cat states~\cite{Zurek:2001aa,Goldberg:2021aa,shukla:2023aa}:
\begin{equation}
\ket{\psi^{(q)}}=\sum_{l=0}^{2q-1}\psi_l\ket{\alpha e^{\tfrac{2\pi\iu l}{2q}}}.
\end{equation}
Each of these states provides the same value $| \langle \hat{T}_{Kq} \rangle |^{2} =\left|\alpha\right|^{4K}$. Then, since saturating the inequality for all $q$ requires $\psi_0=0$, only a coherent state maximizes the cumulative sum  ${A}_M$ for any fixed energy $\bar{n}=|\alpha|^2$.

We already know that $\vac$ minimizes $A_M$. For a given, fixed $\bar{n}>0$, one can ask what state minimizes the cumulative multipoles. All of the multipoles with $q\neq 0$ vanish for Fock states; this is because they vanish for any state that is unchanged after undergoing a rotation by $\pi/2q$ about the origin in phase space. The $q=0$ multipoles, on the other hand, depend only on the diagonal coefficients of the density matrix in the Fock basis, which can be minimized in parallel.

To minimize a multipole moment 
\begin{equation}
| \langle \hat{T}_{K0} \rangle | = K!\sum_{n\geq K}\binom{n}{K}\varrho_{nn},
\end{equation}
there are two cases to consider: $\bar{n}< K$ and $\bar{n}\geq K$. If $\bar{n}< K$, the multipole vanishes by simply partitioning all of the probability among the Fock states with fewer than $K$ photons and arranging those states in a convex combination with no coherences in the Fock basis. If $\bar{n}\geq K$, the sum is ideally minimized by setting $\varrho_{\bar{n}\bar{n}}=1$, by convexity properties of the binomial coefficients (they grow by a larger amount when $n$ increases than the amount that they shrink when $n$ decreases). For noninteger $\bar{n}$, the minimum is achieved by setting 
\begin{equation}
\varrho_{\lceil \bar{n}\rceil\lceil \bar{n}\rceil}=1-(\lceil \bar{n}\rceil-\bar{n}),
\qquad \qquad 
\varrho_{\lceil \bar{n}\rceil-1 \, \lceil \bar{n}\rceil-1}=\lceil \bar{n}\rceil-\bar{n}
\end{equation} 
with no coherences between these two Fock states. Here, $\lceil x \rceil$ is the ceiling function that gives  the smallest integer value that is bigger than or equal to $x$. Since this minimization does not depend on $K$, we have thus found the unique state that minimizes $A_M$ for all $M$ with arbitrary $\bar{n}$:
\begin{equation}
\arg\max A_M(\hat{\rho}|\bar{n})=(\lceil \bar{n}\rceil-\bar{n})\ket{\lceil\bar{n}\rceil-1}\bra{\lceil\bar{n}\rceil-1}
+
(1+\bar{n}-\lceil \bar{n}\rceil)\ket{\lceil\bar{n}\rceil-1}\bra{\lceil\bar{n}\rceil-1}.
\end{equation}
It is intriguing that coherent states and Fock states respectively maximize and minimize this sum for integer-valued energies, while a convex combination of the nearest-integer Fock states minimize this sum for a noninteger energy. These results should be compared against those for the sum $\mathfrak{A}_M$, which was uniquely maximized by the vacuum state that minimizes the sums here and for which the states that made it vanish were Fock states with large energies. Both sums are minimized for some Fock states and both sums are maximized by some coherent states, but the scalings with energy are opposite, where smaller energy leads to larger $\mathfrak{A}_{M}$ and smaller ${A}_{M}$ while larger energy leads to smaller $\mathfrak{A}_{M}$ and larger $A_{M}$; it just so happens that the state with smallest energy is both a Fock state and a coherent state.

As a preview for the next section, we note that states with a finite number of photons $2K$ are described by only the moments up to $K$. To directly reconstruct the state, one simply arranges these measured moments with our inverse operators as in Eq.~\eqref{eq:rho from inverse}; the operators described here are exactly those used to sort measurement information for state reconstruction.

\section{Further applications: quantizers and tomography}
\label{sec:applications}

We have analyzed moments of a quantum state in the monomial basis to gain intuition as to the impact of each moment on the quantumness of the state. What else can be done with these moments?

One direct application is to creating a state's phase-space functions from a set of measurements; this is the province of quantum state tomography. To create any $s$-ordered quasiprobability distribution from a given state $\hat{\varrho}$, one uses the operator kernel $\hat{w}^{(s)}(\alpha)$ to find
\begin{equation}
    W^{(s)}_\varrho(\alpha)=\mathrm{Tr}[\hat{w}^{(s)}(\alpha)\hat{\varrho}].
\end{equation} The Wigner function, for example, is found by taking $s=0$, while $s$ can range from $1$ for the normally ordered quasiprobability distribution to $-1$ for the antinormally ordered counterpart. Without yet giving a form for the kernels $\hat{w}^{(s)}(\alpha)$, we note that it would be useful to express them in the monomial basis
\eq{
\hat{w}^{(s)}(\alpha)=\sum_{Kq}w_{Kq}^{(s)}(\alpha)\hat{T}_{Kq}.
} Then, a measurement of the moments $\langle \hat{T}_{Kq}\rangle_\varrho$, where we now make explicit that the expectation values are with respect to the state $\hat{\varrho}$,  which is achieved via homodyne detection as above, immediately yields the quasiprobability distribution through
\begin{equation}
    W^{(s)}_\varrho(\alpha)=\sum_{Kq}w_{Kq}^{(s)}(\alpha)\langle \hat{T}_{Kq}\rangle_\varrho.
\end{equation} The missing connection between the tomographic measurement results $\langle \hat{T}_{Kq}\rangle_\varrho$ and the quasiprobability distributions $W^{(s)}_\varrho(\alpha)$ are the values of the coefficients $w_{Kq}^{(s)}(\alpha)$. Our formalism directly solves this problem: they are found via our inverse operators
\begin{equation}
    w_{Kq}^{(s)}(\alpha)=\mathrm{Tr}[\hat{w}^{(s)}(\alpha) \hat{\mathfrak{T}}_{Kq}].
\end{equation}

We can immediately compute the coefficients of the operator kernels in the monomial basis from the definitions
\begin{equation}
    \hat{w}^{(s)}(\alpha)=\frac{1}{\pi}\int d^2\beta e^{s\tfrac{|\beta|^2}{2}+\alpha\beta^*-\alpha^*\beta} \hat{D}(\beta).
\end{equation} 
The Wigner function, for example, has $\hat{w}^{(0)}(\alpha)$ and we have already computed $\mathrm{Tr}[\hat{D}(\beta)\hat{\mathfrak{T}}_{Kq}]$ in Eq.~\eqref{eq:Tr D inv}. This leaves us with one integral to solve:
\begin{align}
        w_{Kq}^{(0)}(\alpha)&=\frac{(-1)^{2K}}{\pi}\int d^2\beta \eu^{\alpha\beta^*-\alpha^*\beta-\tfrac{|\beta|^2}{2}}\frac{\beta^{K+q}(-\beta^*)^{K-q}}{(K+q)!(K-q)!} \nonumber \\
        &=(-1)^{2q}\frac{2^{K+1}\eu^{-2|\alpha|^2}}{(K+q)!}(\sqrt{2}\alpha)^{2q}L_{K-q}^{2q}(2|\alpha|^2) \nonumber \\
        &=\frac{2^{K+1}\eu^{-2|\alpha|^2}}{(K-q)!}(\sqrt{2}\alpha^*)^{-2q}L_{K+q}^{-2q}(2|\alpha|^2).
 \end{align}
The same calculation can be performed with any other value of $s$ by simply replacing the argument of the exponent in the quantizer. As well, a tomography scheme that directly measures moments with another ordering, such as the symmetrically ordered moments $\langle \hat{T}^W_{Kq}\rangle$ defined in App.~\ref{sec:symord}, can again be used to directly construct any quasiprobability distribution after computing the moments of our appropriate inverse operator, such as $\mathrm{Tr}[\hat{w}^{(s)}(\alpha)\hat{\mathfrak{T}}^W_{Kq}]$. We list some relevant results along these lines:
\begin{eqnarray}
    \Tr [\hat{w}^{(1)}(\alpha) \hat{\mathfrak{T}}_{Kq}]&=&\Tr[\hat{w}^{(0)}(\alpha)\hat{\mathfrak{T}}^W_{Kq}]=\frac{(-1)^{K-q}}{(K+q)!(K-q)!}\frac{\partial ^{2K}}{\partial^{K+q}\alpha\partial^{K-q}\alpha^*}\delta^2(\alpha),
    \label{eq:quantizer s1 deltas}
    \\
    \Tr [\hat{w}^{(s)}(\alpha)\hat{\mathfrak{T}}_{Kq}]&=&\Tr[\hat{w}^{(s-1)}(\alpha)\hat{\mathfrak{T}}_{Kq}^W]=(1-s)^{-(K+1)}w_{Kq}^{(0)}(\alpha/\sqrt{1-s}),
    \label{eq:quantizer moment any s}
    \\
    \Tr [\hat{w}^{(0)}(\alpha)\hat{\mathfrak{T}}_{Kq}]&=& \Tr[\hat{w}^{(-1)}(\alpha)\hat{\mathfrak{T}}^W_{Kq}]=w_{Kq}^{(0)},\\
    \Tr[\hat{w}^{(-1)}(\alpha)\hat{\mathfrak{T}}_{Kq}]&=&\frac{e^{-|\alpha|^2}}{(K-q)!}(\alpha^*)^{-2q}L_{K+q}^{-2q}(|\alpha|^2).
\end{eqnarray}

The inverse operators and multipole moments are thus intimately connected to quantizers and tomography. It may come as no surprise that these are even more intertwined: a little inspection leads to
\begin{equation}
\begin{aligned}
    \hat{w}^{(-1)}(\alpha)& = \sum_{Kq}\hat{\mathfrak{T}}_{Kq}\alpha^{*K+q}\alpha^{K-q}
=\sum_{Kq}\hat{\mathfrak{T}}_{Kq}\langle \alpha|\hat{T}_{Kq}|\alpha\rangle \\
   \hat{w}^{(0)}(\alpha) &=\sum_{Kq}\hat{\mathfrak{T}}_{Kq}^W\langle \alpha|\hat{T}_{Kq}|\alpha\rangle,
   \end{aligned}
\end{equation}
 and so on for any $s$:
\begin{equation}
    \hat{w}^{(-s)}(\alpha)
    =\frac{1}{\pi}\int d^2\beta e^{-s\tfrac{|\beta|^2}{2}}\hat{D}(\beta)\sum_{Kq}\frac{(\alpha\beta^*)^{K-q}}{(K-q)!}\frac{(-\alpha^*\beta)^{K+q}}{(K+q)!}
    =\sum_{Kq}\hat{\mathfrak{T}}_{Kq}^{(s)}\langle \alpha|\hat{T}_{Kq}|\alpha\rangle,
\end{equation} with the $s$-ordered inverse operators
\begin{equation}
    \hat{\mathfrak{T}}_{Kq}^{(s)}=
    \frac{\left(-1\right)^{K+q}}{\left(K+q\right)!\left(K-q\right)!} \frac{1}{\pi} \int d\beta^2 \; e^{-s\tfrac{|\beta|^2}{2}}\hat{D} (\beta ) \; \beta^{K+q}\beta^{\ast K-q}.
\end{equation}
The monomial moments for a coherent state are the expansion coefficients of the $(-s)$-ordered kernels in the $s$-ordered inverse operator basis and the quasiprobability distributions can now be written as
\begin{equation}
    W^{(s)}_\varrho(\alpha)=\sum_{Kq}\langle \hat{\mathfrak{T}}_{Kq}^{(-s)}\rangle_\varrho\langle \alpha |\hat{T}_{Kq}|\alpha\rangle
    =\sum_{Kq}\langle \hat{\mathfrak{T}}_{Kq}^{(-s)}\rangle_\varrho\alpha^{* K+q}\alpha^{K-q}
    ,
\end{equation} where $\hat{\mathfrak{T}}_{Kq}^{(0)}=\hat{\mathfrak{T}}_{Kq}^{W}$ and $\hat{\mathfrak{T}}_{Kq}^{(1)}=\hat{\mathfrak{T}}_{Kq}$ using the notation from before. The multipole moments are exactly what are required to be measured for constructing a quasiprobability distribution in the polynomial (i.e., $\alpha^{* K+q}\alpha^{K-q}$) basis. Then, different polynomial bases can be obtained by switching $\hat{T}_{Kq}$ for a different ordering in these expressions.

As a sort of duality, we can compute
    \begin{align}
        \sum_{Kq}\langle \alpha |\hat{\mathfrak{T}}_{Kq}^{(s)}|\alpha\rangle  \hat{T}_{Kq}
    &=\frac{1}{\pi}\int d^2\beta \langle \alpha|\hat{D}(\beta)|\alpha\rangle e^{-s\tfrac{|\beta|^2}{2}}e^{-\beta\had}e^{\beta^*\ha } \nonumber 
    \\
    &=\frac{1}{\pi}\int d^2\beta \langle \alpha|\hat{D}(\beta)|\alpha\rangle \hat{D}(-\beta)e^{-s\tfrac{|\beta|^2}{2}+\tfrac{|\beta|^2}{2}} \nonumber \\
    &=\frac{1}{\pi}\int d^2\beta e^{-s\tfrac{|\beta|^2}{2}+\alpha\beta^*-\alpha^*\beta} \hat{D}(\beta) \nonumber \\
    &=\hat{w}^{(-s)}(\alpha) \, .
    \end{align}
For example, we have found the quantizer for the Husimi $Q$ function to be
\begin{equation}
    \hat{w}^{(-1)}(\alpha)=\sum_{Kq}\langle \alpha |\hat{\mathfrak{T}}_{Kq}^{(1)}|\alpha\rangle  \hat{T}_{Kq},
\end{equation}
such that
\begin{equation}
    Q_\varrho(\alpha)=\Tr[\hat{w}^{(-1)}(\alpha)\hat{\varrho}]=\langle\alpha|\hat{\varrho}|\alpha\rangle;
\end{equation}
this is a specific case of the general result from orthonormality:
\begin{equation}
    |\langle \psi|\phi\rangle|^2=\sum_{Kq}\langle \psi|\hat{\mathfrak{T}}_{Kq}|\psi\rangle\langle \phi |\hat{T}_{Kq}|\phi\rangle.
\end{equation}
These lead to the symmetrically pleasing results that intertwine the monomials and their inverses:
\begin{eqnarray}
    \hat{w}^{(s)}(\alpha)& = & \sum_{Kq} \hat{\mathfrak{T}}_{Kq}^{(-s)} \langle \alpha|\hat{T}_{Kq}|\alpha\rangle=\sum_{Kq}\langle \alpha|\hat{\mathfrak{T}}_{Kq}^{(-s)}|\alpha\rangle  \hat{T}_{Kq};\\
    W^{(s)}_\varrho(\alpha) & = & \sum_{Kq}\langle \hat{\mathfrak{T}}_{Kq}^{(-s)}\rangle_\varrho \langle \alpha|\hat{T}_{Kq}|\alpha\rangle=\sum_{Kq}\langle \alpha|\hat{\mathfrak{T}}_{Kq}^{(-s)}|\alpha\rangle \langle \hat{T}_{Kq}\rangle_\varrho.
\end{eqnarray}
To summarize, a state's quasiprobability distributions can be found either by measuring the monomials and then weighting those measurements by the $s$-ordered moments of coherent states, or by measuring the $s$-ordered moments and weighting those measurements by the inverse moments (monomial expectation values) of coherent states.

None of the operators $\hat{T}_{Kq}$ or $\hat{\mathfrak{T}}_{Kq}$ are trace class, with formally infinite trace when $q=0$ and vanishing trace when $q\neq 0$. Yet, by matching all of these operators together, another connection we can write along these lines is
\begin{equation}
    \sum_{Kq} \hat{\mathfrak{T}}_{Kq}\hat{T}_{Kq}=\frac{1}{\pi}\int d^2\beta \hat{D}(\beta)e^{-\tfrac{|\beta|^2}{2}}e^{-\beta \had}e^{\beta^*\ha}=\frac{1}{\pi}\int d^2\beta \hat{D}(\beta)\hat{D}(-\beta)=\hat{\openone}\frac{1}{\pi}\int d^2\beta.
\end{equation} The sum certainly does not converge in any usual sense, as could have been expected for identity operators in infinite dimensions, but the sum of the operators paired with their inverse operators can conclusively be asserted to be proportional to this identity operator.

Finally, on the topic of operator ordering, the question frequently arises: how does one actually express some $s$-ordered polynomial operator in a known basis, such as the normally ordered one? Our construction is the direct solution: the expansion coefficients are given by the trace of the product of the polynomial with our inverse operators $\hat{\mathfrak{T}}_{Kq}$, most of which can simply be read off from Eq.~\eqref{eq:quantizer moment any s} [other than $s=1$, which typically require derivatives of delta functions as in Eq.~\eqref{eq:quantizer s1 deltas}]. These fill a missing gap and can readily be used for further computations.

To put it all together, let us consider a measurement of the $s$-ordered polynomials. We wish to construct an arbitrary $s^\prime$-ordered quasiprobabilty distribution. This is achieved via
\begin{eqnarray}
    \hat{w}^{(s^\prime)}(\alpha) & = & \sum_{Kq}\langle \alpha|\hat{\mathfrak{T}}_{Kq}^{(s-s^\prime-1)}|\alpha\rangle\hat{T}^{(s)}_{Kq}=\sum_{Kq}\hat{\mathfrak{T}}_{Kq}^{(s-s^\prime-1)}\langle \alpha|\hat{T}^{(s)}_{Kq}|\alpha\rangle,\\
    W_\varrho^{(s^\prime)}(\alpha) & = & \sum_{Kq}\langle \alpha|\hat{\mathfrak{T}}_{Kq}^{(s-s^\prime-1)}|\alpha\rangle\langle \hat{T}^{(s)}_{Kq}\rangle_\varrho= \sum_{Kq}\langle \hat{\mathfrak{T}}_{Kq}^{(s-s^\prime-1)}\rangle_\varrho \langle \alpha|\hat{T}^{(s)}_{Kq}|\alpha\rangle
\end{eqnarray}
for any $s$-ordered polynomials 
\begin{equation}
\hat{T}_{Kq}^{(s)}=\tfrac{\partial^{2K}}{\partial \alpha^{K+q}\partial(-\alpha^*)^{K-q}}\hat{D}(\alpha)\eu^{s|\alpha|^2/2}\big|_{\alpha=0} 
\end{equation} that arise from noise-added channels that perform quantum-limited amplification and then attenuation on a state by the same factor $2/(1+s)$ (e.g., $\had\ha\to \had\ha+(1-s)/2$; equivalently, the quasiprobability distribution for the operators get smoothened by a Gaussian kernel for increasing noise).
A homodyne detection scheme that measures the symmetric polynomials $\langle \hat{T}^{W}_{Kq}\rangle_\varrho=\langle \hat{T}^{(0)}_{Kq}\rangle_\varrho$, for example, yields any $s^\prime$-ordered quasiprobability distribution by simply summing the measurements with weights
$\langle \alpha|\hat{\mathfrak{T}}_{Kq}^{(-s^\prime-1)}|\alpha\rangle$.

\section{Concluding remarks}
\label{sec:conc}

Expanding the density operator in a conveniently chosen operator set has considerable advantages. By  using explicitly the algebraic properties of the basis operators the calculations are often greatly simplified. But the usefulness of the method depends on the choice of the basis operator set. The idea of irreducible tensor operators is to provide a well-developed and efficient way of using the inherent symmetry of the system. 

However, the irreducible-tensor machinery was missing for CV, in spite of the importance of these systems in modern quantum science and technology. We have provided a complete account of the use of such bases, which should constitute an invaluable tool for quantum optics.

\section*{Acknowledgments}
We thank H. de Guise and U. Seyfarth for discussions. This work received funding from the European Union’s Horizon 2020 research and innovation programme project STORMYTUNE under grant Agreement No. 899587. AZG acknowledges that the NRC headquarters is located on the traditional unceded territory of the Algonquin Anishinaabe and Mohawk people, as well as support from the NSERC PDF program. LLSS acknowledges support from Ministerio de Ciencia e Innovaci\'on (Grant  PID2021-127781NB-I00).

\appendix
\section{Transformation properties of the operators}
\label{sec:apen}
We present in this appendix some properties of the composition law of two tensors operators.  Writing the inverse operators $\hat{\mathfrak{T}}_{Kq}$ in the basis of monomial operators $\hat{T}_{Kq}$ is as simple as reading off coefficients using Fig.~\ref{fig:parts_of_matrix}. We have already identified that each inverse operator $\hat{\mathfrak{T}}_{Kq}$ has contributions from a finite stripe with $K-|q|$ elements along the $q$th diagonal. The monomials, on the other hand, have contributions on the $-q$th stripe, starting from the $(K-|q|)$th element and going to infinity. The expansion is thus given by a sum of monomials $\hat{T}_{K-q}$ for all possible values of $K$ up until infinity, whose expansion coefficients can be found iteratively. The coefficient with the lowest value of $K$ is just given by the coefficient of the top-left element of $\hat{\mathfrak{T}}_{Kq}$ in Fig.~\ref{fig:parts_of_matrix}. The coefficient with the next-lowest value of $K$ can be found iteratively by canceling the contribution from the monomial that begins at the top-left corner and adding the contribution from the monomial that begins after the top-left corner. The iteration must continue to infinity in order to make sure all of the contributions after the $(2K+1)$th antidiagonal vanish.

Another method of finding these expansion coefficients considers the quantity $\Tr(\hat{\mathfrak{T}}_{Kq}\hat{\mathfrak{T}}_{K^\prime q^\prime})$. We already know by inspection that this will vanish unless $q=-q^\prime$. We can directly compute these overlaps by summing terms from Eq.\eqref{eq:inverse operators Fock basis}:
\begin{equation}
\begin{aligned}
\hat{\mathfrak{T}}_{Kq} & = \sum_{K^\prime q^\prime }\hat{T}_{K^\prime q^\prime}\Tr(\hat{\mathfrak{T}}_{Kq}\hat{\mathfrak{T}}_{K^\prime q^\prime})  \\
\Tr(\hat{\mathfrak{T}}_{Kq}\hat{\mathfrak{T}}_{K^\prime q^\prime})&=\delta_{q,-q^\prime}\frac{(-1)^{K+K^\prime+2 |q|} \, _2F_1(|q|-K,|q|-K^\prime;2 |q|+1;1)}{(2 |q|)! (K-|q|)! (K^\prime-|q|)!},
\end{aligned}
\end{equation}
which provides a useful alternative formula for the integrals
\begin{align}
\Tr(\hat{\mathfrak{T}}_{Kq}\hat{\mathfrak{T}}_{K^\prime q^\prime})& =
\frac{\left(-1\right)^{K+K^\prime+q+q^\prime}}{\left(K+q\right)!\left(K-q\right)!\left(K^\prime+q^\prime\right)!\left(K^\prime-q^\prime\right)!} \nonumber \\
&\times \frac{1}{\pi^{3}} \int d^2\alpha d^2\beta d^2\gamma e^{-\tfrac{|\beta|^2+|\gamma|^2}{2}}\bra{\alpha}\hat{D}\left(\beta\right)\hat{D}\left(\gamma\right)\ket{\alpha}\beta^{K+q}\beta^{*K-q}\gamma^{K^\prime+q^\prime}\gamma^{*K^\prime-q^\prime}.
\end{align}
Just because a particular product $\hat{\mathfrak{T}}_{Kq}\hat{\mathfrak{T}}_{K^\prime q^\prime}$ with $q\neq q^\prime$ is traceless does not mean that it necessarily vanishes. In fact, we can directly compute the product of two such operators to find their structure constants. Each inverse operator $\hat{\mathfrak{T}}_{Kq}$ serves to decrease the number of photons in a state by $2q$, so the product of two inverse operators must be a finite sum of inverse operators whose second index satisfies $q^{\prime\prime}=q+q^\prime$.

We start by writing
\begin{equation}
\hat{\mathfrak{T}}_{Kq}\hat{\mathfrak{T}}_{K^\prime q^\prime}=\sum_{K^{\prime\prime}}f_{K^{\prime\prime}}(K,K^\prime,q,q^\prime) \hat{\mathfrak{T}}_{K^{\prime\prime},q+q^\prime}.
\end{equation} 
In theory, the coefficients $f_{K^{\prime\prime}}$ are formally given by $\Tr(\hat{\mathfrak{T}}_{Kq}\hat{\mathfrak{T}}_{K^\prime q^\prime}\hat{T}_{K^{\prime\prime},q+q^\prime})$. Inspecting Eq.~\eqref{eq:inverse operators Fock basis}, we find some interesting, immediate results: for example, when $q,q^\prime\geq 0$ and $2q>K^\prime -q^\prime$, all of the structure constants $f_{K^{\prime\prime}}$ vanish and we have $\hat{\mathfrak{T}}_{Kq}\hat{\mathfrak{T}}_{K^\prime q^\prime}=0$. Similar vanishing segments can be found for any combination of the signs of $q$ and $q^\prime$, which is not readily apparent from multiplications of displacement operators from Eq.~\eqref{eq:inverse operators integrals}.

The nonzero structure constants can be found via iteration, using Fig.~\ref{fig:parts_of_matrix} as a guide. Taking, for example, $q,q^\prime \geq 0$, 
we find products of the form
\begin{equation}
\hat{\mathfrak{T}}_{Kq}\hat{\mathfrak{T}}_{K^\prime q^\prime}=\sum_{n=2q+2q^\prime}^{\min(K^\prime+q^\prime,K+q+2q^\prime)} \! \! \! \! \! \frac{(-1)^{K+K^\prime+q-q^\prime}(K+q+2q^\prime-n)!^{-1}}{\sqrt{n!(n-2q-2q^\prime)!}(n-2q^\prime)!(K^\prime+q^\prime-n)!} \ket{n-2q-2q^\prime}\bra{n} ;
\end{equation}
the nonzero structure constants obey $K^{\prime\prime}\leq K^{\prime\prime}_{\mathrm{max}}= \min(K+q^\prime,K^\prime-q)$. The one with the largest $K^{\prime\prime}$ is the only one that has the term $\ket{K^{\prime\prime}_{\mathrm{max}}-q-q^\prime}\bra{K^{\prime\prime}_{\mathrm{max}}+q+q^\prime}$, so its structure constant must balance the unique contribution to that term from $\hat{\mathfrak{T}}_{K^{\prime\prime}_{\mathrm{max}},q+q^\prime}$. This means that
\begin{equation}
f_{K^{\prime\prime}_{\mathrm{max}}}(K,K^\prime,q,q^\prime)
=\frac{(-1)^{K+K^\prime+q-q^\prime}}
{(K^{\prime\prime}_{\mathrm{max}}+q-q^\prime)!(K^\prime-q-K^{\prime\prime}_{\mathrm{max}})!(K+q^\prime-K^{\prime\prime}_{\mathrm{max}})!},
\end{equation} 
where one of the final two terms in the denominator will simply be $0!=1$. Then, by iteration, one can balance the contribution of $\hat{\mathfrak{T}}_{K^{\prime\prime}_{\mathrm{max}}-k,q+q^\prime}$ in order to find the structure constants $f_{K^{\prime\prime}_{\mathrm{max}}-k}(K,K^\prime,q,q^\prime)$.

The structure constants for the monomial operators are already known. One can compute~\cite{Gosson:2016aa} 
\begin{equation}
\hat{T}_{Kq}\hat{T}_{K^\prime q^\prime}=\sum_n c_n \had^{K+q+K^\prime+q^\prime-n} \ha^{K+K^\prime-q-q^\prime-n}
\end{equation}
from normal ordering. 

The inverse operators transform nicely under displacements:
\begin{align}
\hat{D}(\alpha) \hat{\mathfrak{T}}_{Kq} \hat{D}(\alpha)^\dagger &= \frac{(-1)^{K+q}}{\pi (K+q)! (K-q)!} \int d^2\beta e^{|\beta|^2/2} e^{\alpha \beta^{\ast} - \alpha^{\ast} \beta} \hat{D}(\beta ) \beta^{K+q}\beta^{\ast K-q} \nonumber \\
&=
\sum_{S}^\infty \sum_{l=-S}^S\alpha^{S-l}\alpha^{\ast S+l}\binom{K+S+q+l}{K+q}\binom{K+S-q-l}{K-q}
\hat{\mathfrak{T}}_{K+S,q+l}.
\end{align}
These displaced operators are inverse to the displaced monomials \eq{
\hat{D}(\alpha)\hat{T}_{Kq}\hat{D}(\alpha)^\dagger=\sum_{S=0,1/2}^K\sum_{l=-S}^S\binom{K+q}{S+l}\binom{K-q}{S-l}(-\alpha^*)^{K+q-S-l}(-\alpha)^{K-q-S+l}\hat{T}_{Sl}.
} It is interesting to note that the displaced inverse operators are given by an infinite sum of inverse operators and the displaced monomials by a finite sum of monomials, in contrast to the number of terms $\ket{m}\bra{n}$ required to expand the original operators in the Fock basis.

\section{Symmetrically ordered monomials}
\label{sec:symord}

We briefly consider here the example of symmetrically ordered multinomials $\hat{T}_{Kq}^W$. We can write them explicitly in terms of the normally ordered polynomials as
\begin{equation}
\hat{T}_{Kq}^W=\{\had^{K+q}\ha^{K-q}\}_{\mathrm{sym}}=\sum_{n=0}^{\min(K+q,K-q)}\frac{(K+q)!(K-q)!}{2^n n!(K+q-n)!(K-q-n)!}\hat{T}_{K-n,q} \, ,
\end{equation}
where $\{ \cdot \}_{\mathrm{sym}}$ denotes the symmetric (or Weyl) order or operators~\cite{Gosson:2016aa}.  An important expression for the symmetrically ordered polynomials is
\begin{equation}
\hat{T}_{Kq}^W=\frac{\partial^{2K}}{\partial\beta^{K+q}\partial(-\beta^*)^{K-q}}\hat{D}(\beta)\bigg|_{\beta=0}.
\end{equation} 
We thus look for inverse operators through
\begin{align}
\Tr (\hat{\mathfrak{T}}_{Kq}^W\hat{T}_{K^\prime q^\prime}^W) & =
\frac{\partial^{2K^\prime}}{\partial\beta^{K^\prime+q^\prime}\partial(-\beta^\ast)^{K^\prime-q^\prime}}\Tr [\hat{\mathfrak{T}}_{Kq}^W\hat{D}(\beta)]\bigg|_{\beta=0} \nonumber  \\
& = \frac{1}{\pi}\int {d^2\beta} \Tr [\hat{D}(-\beta) \hat{\mathfrak{T}}_{Kq}^W ]\Tr\left[\hat{D}(\beta)\frac{\partial^{2K}}{\partial\alpha^{K+q}\partial(-\alpha^*)^{K-q}}\hat{D}(\alpha)\bigg|_{\alpha=0}\right]\nonumber \\
& = \int d^2\beta\Tr [\hat{D}(-\beta)\hat{\mathfrak{T}}_{Kq}^W] \; \frac{\partial^{2K}}{\partial\alpha^{K+q}\partial(-\alpha^\ast)^{K-q}} \delta^2(\alpha+\beta)\bigg|_{\alpha=0}.
\end{align} 
By inspection, we attain orthonormality when
\begin{equation}
\Tr [\hat{\mathfrak{T}}_{Kq}^W\hat{D}(\beta) ]=\frac{\beta^{K+q}(-\beta^\ast)^{K-q}}{(K+q)!(K-q)!},
\end{equation} 
which corresponds to
\begin{equation}
\hat{\mathfrak{T}}_{Kq}^W= \frac{1}{\pi} \int {d^2\beta} \hat{D}(-\beta)\frac{\beta^{K+q}(-\beta^\ast)^{K-q}}{(K+q)!(K-q)!}=\frac{(-1)^{K+q}}{\pi (K+q)!(K-q)!}\int {d^2\beta}\hat{D}(\beta)\beta^{K+q}\beta^{\ast K-q},
\end{equation} 
simply differing from the expression \eqref{eq:inverse operators integrals} for $\hat{\mathfrak{T}}_{Kq}$ by removing the factor of $\exp(-|\beta|^2/2)$.

We can find the multipoles for specific states.  We simply quote the results
\begin{align}
\langle \alpha | \hat{\mathfrak{T}}_{Kq}^W | \alpha \rangle & = \frac{2^{K-q+1}(-1)^{K+q}}{(K-q)!}\frac{\eu^{-2|\alpha|^2}}{\alpha^{\ast 2q}}L_{K+q}^{(-2q)}(2|\alpha|^2)
\label{eq:Wigner multipoles coherent state}
\end{align} 
and
\begin{align}
\langle n | \hat{\mathfrak{T}}_{Kq}^W | n \rangle & =\delta_{q0}\frac{(-1)^K}{K!^2}2\int_0^\infty r^{2K+1}\eu^{-r^2/2}L_n(r^2)=\delta_{q0}\frac{(-1)^K 2^{K+1} \, _2F_1(K+1,-n;1;2)}{K!}.
\end{align} 
For arbitrary states, we can follow the same procedure as we used for normal order; the final result is   ($m\leq n$)
\begin{align}
\langle n | \hat{\mathfrak{T}}_{Kq}^W | m \rangle &=\frac{\left(-1\right)^{K+q}}{\left(K+q\right)!\left(K-q\right)!}\int \frac{d^2\beta}{\pi} \eu^{-|\beta|^2/2}\sqrt{\frac{n!}{m!}}\beta^{m-n}L_n^{(m-n)}(|\beta|^2)\beta^{K+q}\beta^{*K-q} \nonumber \\
&=\delta_{n-m \; 2q}\frac{(-1 )^{K+3q}}{(K+q)!}\sqrt{\frac{n!}{(n-2q)!}}  2^{K+q+1} \, _2\tilde{F}_1(k+q+1,2 q-n;2 q+1;2) .
\end{align} 

Finally, it is direct to check that the tensors $\hat{T}_{Kq}^{W}$ are covariant under symplectic transformations~\cite{Ivan:2012aa}.

\section{Vacuum state as maximizing the cumulative multipolar distribution}
\label{app:vac max}

We here provide analytical and numerical evidence that the vacuum state uniquely maximizes the cumulative multipolar distribution to arbitrary orders $M>3/2$.

First, we note by convexity that the multipole moments are all largest for pure states.
We next ask how to maximize a single multipole moment $|\langle \hat{\mathfrak{T}}_{Kq} \rangle|$. The phases can be arranged such that $\varrho_{nm}(-1)^n > 0$ for all $n$ and $m$ in Eq.~\eqref{eq:multipole moments in terms of fock basis}, while each term is bounded as $|\varrho_{nm}|\leq\sqrt{\varrho_{mm}\varrho_{nn}}$. It is tempting to use a Cauchy-Schwarz inequality to say that this expression is maximized by states with the relationship $\varrho_{nn}=\lambda n!$ for some normalization constant $\lambda$. This fails, however, for two related reasons: one cannot simultaneously saturate the inequality $|\varrho_{nm}|\leq\sqrt{\varrho_{mm}\varrho_{nn}}$ for all $m$ and $n$ while retaining a positive density operator $\hat{\varrho}$; similarly, the trace of $\hat{\varrho}$ is bounded, which the Cauchy-Schwarz inequality does not take into consideration. One can outperform this Cauchy-Schwarz bound by concentrating all of the probability in the term with the largest value of $1/\sqrt{n!(n-2q)!(K+q-n)!^2}$. Taking 
\begin{equation}
\tilde{n}=\arg\max_n\frac{1}{\sqrt{n!(n-2q)!}(K+q-n)!,}
\end{equation} 
$| \langle \hat{\mathfrak{T}}_{Kq} \rangle | $ is maximized by any pure state with $\varrho_{\tilde{n}\tilde{n}}=\varrho_{\tilde{n}-2q,\tilde{n}-2q}=1/2$:
\begin{equation}
\max | \langle \hat{\mathfrak{T}}_{Kq} \rangle |^2=\frac{1}{4\tilde{n}!(\tilde{n}-2)q!(K+q-\tilde{n})!^2}
\label{eq:max individual multipole}
\end{equation}
This condition changes with $K$ and $q$, so there will always be a competition between which terms $| \langle \hat{\mathfrak{T}}_{Kq} \rangle|^2$ are maximized in the cumulative sum.

The contributions to $\mathfrak{A}_M$ by the various terms $| \langle \hat{\mathfrak{T}}_{Kq} \rangle |^2$ diminish with increasing $K$, which can be seen through the following argument. As $M$ increases by $1/2$, the number of new terms contributing to the sum increases quadratically: there are $2M+1$ new multipoles  to consider and each multipole is a sum of at most $M+1$ terms. From the preceding discussion, each multipole is individually maximized when it is made from only a single term, the cumulative multipole moment $\mathfrak{A}_M$ can only increase by the addition of $\mathcal{O}(M)$ (competing) terms. In contrast, the magnitudes of each of the multipole moments decay exponentially with increasing $M$, due to the factorials in the denominator Eq.~\eqref{eq:max individual multipole}, stemming from Eq.~\eqref{eq:multipole moments in terms of fock basis}. One can, therefore, guarantee that a state maximizing $\mathfrak{A}_M$ for sufficiently large $M$ will continue to maximize $\mathfrak{A}_M$ for all larger values of $M$, at least approximately.

\begin{figure}
    \centering
    \includegraphics{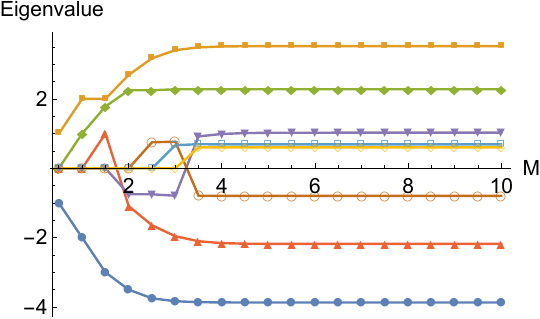}
    \caption{Eigenvalues of $\hat{\mathfrak{A}}_M$ with the eight largest magnitudes up until $M=10$. The negative eigenvalue with the largest magnitude corresponds to the entangled state $\ket{0}\otimes\ket{1}-\ket{1}\otimes\ket{2}$, the positive eigenvalue with the largest magnitude is $\ket{0}\otimes\ket{2}-c\ket{1}\otimes\ket{1}+\ket{2}\otimes\ket{0}$ for some positive constant $c>1$, and the positive eigenvalue with the second largest magnitude is $\ket{0}\otimes\ket{0}$. These dictate that the symmetric state $\ket{\psi}\otimes\ket{\psi}$ for which the expectation value of $\hat{\mathfrak{A}}_M$ is largest must be confined to the sector spanned by $\ket{0}$, $\ket{1}$, and $\ket{2}$.}
    \label{fig:eigenvalues converging}
\end{figure}

We can also inspect the inverse operators directly to understand the maximization properties.
The multipoles being summed as an indicator of quantumness, $| \langle \hat{\mathfrak{T}}_{Kq} \rangle |^2$, can be expressed as expectation values of the duplicated operator $\hat{\mathfrak{T}}_{Kq}\otimes\hat{\mathfrak{T}}_{Kq}^\dagger=\hat{\mathfrak{T}}_{Kq}\otimes\hat{\mathfrak{T}}_{K,-q}$ with respect to the duplicated states $\hat{\varrho}\otimes\hat{\varrho}$. The vacuum state $\ket{0}\otimes \ket{0}$ is the only duplicated state that is an eigenstate of all of the duplicated operators for all $K$ and $q$, albeit with different eigenvalues for each operator. These operators act on Fock states as
\begin{equation}
(\hat{\mathfrak{T}}_{Kq}\otimes\hat{\mathfrak{T}}_{Kq}^\dagger) \ket{n}\otimes\ket{n}\propto\ket{n-2q}\otimes\ket{n+2q}
\end{equation}
and have nonzero matrix elements given by Kronecker products of the stripes found in Fig.~\ref{fig:parts_of_matrix} (some combinations of $K$ $q$, and $n$ cause the proportionality constant to be zero). These can be used to help finding the eigenstates and eigenvalues of the summed joint operators 
\begin{equation}
\hat{\mathfrak{A}}_M=\sum_{K=0}^M \sum_{q=-K}^K \hat{\mathfrak{T}}_{Kq}\otimes\hat{\mathfrak{T}}_{Kq}^\dagger.
\end{equation}
As mentioned previously, each individual operator $\hat{\mathfrak{T}}_{Kq}$ only has null eigenstates, unless $q=0$; this can be seen from the striped pattern in Fig. \ref{fig:parts_of_matrix}. The same is true of the joint operators $\hat{\mathfrak{T}}_{Kq}\otimes\hat{\mathfrak{T}}_{Kq}^\dagger$, but is not true of the summed joint operators $\hat{\mathfrak{A}}_M$. The latter are represented in the Fock basis by sparse antitriangular matrices, which can be visualized by Kronecker products of pairs of matrices from Fig. \ref{fig:parts_of_matrix}. The eigenstates and eigenvalues can thus be found directly for any $M$. For example, the joint Fock state with maximal eigenvalue is the joint vacuum state $\ket{0}\otimes\ket{0}$.

The cumulative operators $\hat{\mathfrak{A}}_M$ have positive expectation values when taken with respect to any duplicated state $\hat{\rho}\otimes\hat{\rho}$. However, $\hat{\mathfrak{A}}_M$ may have negative eigenvalues, because some of the eigenstates may not be of the form $\hat{\varrho}\otimes\hat{\varrho}$. For example, the eigenstate whose eigenvalue has the largest magnitude is always found to be the maximally entangled state $(\ket{0}\otimes\ket{1}-\ket{1}\otimes \ket{0})/\sqrt{2}$, with a large, negative eigenvalue. This is orthogonal to any duplicated state $\hat{\varrho}\otimes\hat{\varrho}$ because the latter is permutation symmetric, not antisymmetric, so we can readily ignore all contributions to $\hat{\mathfrak{A}}_M$ from this part of its spectrum.

Another entangled state is the eigenstate with the next largest eigenvalue: $(\ket{0}\otimes\ket{2}-c\ket{1}\otimes\ket{1}+\ket{2}\otimes\ket{0})/\mathcal{N}$ for some positive constants $c$ and $\mathcal{N}=\sqrt{2+c^2}$. This eigenstate obeys permutation symmetry, so it will contribute to the multipole moments. The maximum contribution will come from a state of the form 
\begin{equation}
\ket{\psi}=\sqrt{p_0}\ket{0}+\sqrt{p_1}\eu^{\iu\psi}\ket{1}-\sqrt{1-p_0-p_1}\ket{2},
\end{equation} 
specifically with $p_0=1-p_0-p_1$. Since $c>1$, the contribution is uniquely maximized by $p_0=0$ and $p_1=1$, so again we need only consider the joint Fock states in the analysis. The overlap of $\ket{1}\otimes \ket{1}$ with this eigenstate is $c^2/\mathcal{N}^2  \approx 0.621$.

\begin{figure}
    \centering
    \includegraphics{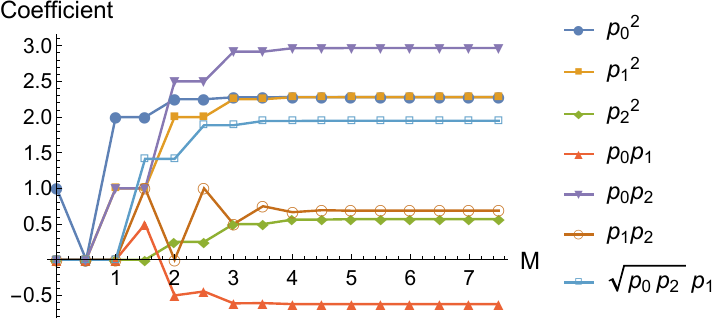}
    \caption{Coefficients of the cumulative mutipole sum for the different weights in the optimal state $\ket{\psi_{\mathrm{opt}}}$. The coefficients rapidly converge for moderate $M$, with those of $p_0^2$ and $p_1^2$ rapidly approaching each other.}
    \label{fig:coefficients converging}
\end{figure}

The eigenstate with the third largest-magnitude eigenvalue is the joint vacuum state $\ket{0}\otimes\ket{0}$. The ratio of its eigenvalue to that with the second largest magnitude approaches $\approx 0.647> c^2/\mathcal{N}^2$ as $M$ increases. This is enough to ensure that the joint vacuum state uniquely maximizes the cumulative multipole moments for all $M$. We stress that these optima have not been found through a numerical optimization, but rather through an exact diagonalization of the operators $\hat{\mathfrak{A}}_M$, which means our analysis does not have to worry about local minima or other numerical optimization hazards.

How can this be made more rigorous? The eigenvalues and eigenstates can be found exactly for any value of $M$ by diagonalizing the sparse matrix $\hat{\mathfrak{A}}_M$. By  $M=9/2$, the largest eigenvalues have already converged to three significant digits and $c^2/\mathcal{N}^2$ to four; by $M=7$, the they have all converged to six significant digits. The contributions from a new, larger value of $K=M$ strictly reduce the magnitude of each expansion coefficient in the sum of Eq. \eqref{eq:inverse operators Fock basis} by a multiplicative factor, ranging from $1/(M+q)$ for the term with the smallest $n$ that has appeared the most times in the cumulative multipole to $1$ for the term with the largest $n$ that has only appeared once previously. There is also the addition of an extra term for $\ket{M-q}\bra{M+q}$, normalized by the large factor $1/\sqrt{(M+q)!(M-q)!}$. Each term gets divided by an increasingly large factor as $M$ increases; the factor that decreases the slowest has already started out with a tiny magnitude due to the normalization factor $1/\sqrt{(M+q)!(M-q)!}$. The magnitudes of the expansion coefficients in the cumulative sums decrease at least exponentially in $\hat{\mathfrak{A}}_{M}-\hat{\mathfrak{A}}_{M-1/2}$, so the largest eigenvalues and eigenstates of $\mathfrak{A}_M$ are fixed once they are known for moderate $M$ (see visualization in Fig.~\ref{fig:eigenvalues converging}).

The above demonstrates that the state maximizing the cumulative multipole moments for any value of $M$ must take the form ($p_0+p_1+p_2=1$)
\begin{equation}
\ket{\psi_{\mathrm{opt}}}=\sqrt{p_0}\ket{0}+\sqrt{p_1}\eu^{\iu\psi}\ket{1}+\sqrt{p_2}\eu^{\iu\phi}\ket{2},
\end{equation} 
because such a states concentrates maximal probability in the subspace with the largest eigenvalues of $\hat{\mathfrak{A}}_M$. We can compute the cumulative multipole moments for such a state, which equal
\begin{align}
\mathfrak{A}_M (\ket{\psi_{\mathrm{opt}}}) & =\sum_{K\in\mathbb{Z}}^M \left|\frac{\varrho_{00}}{K!}-\frac{\varrho_{11}}{(K-1)!}+\frac{\varrho_{22}}{2!(K-2)!}\right|^2+2\frac{\left|\varrho_{20}\right|^2}{2(K-1)!^2} \nonumber \\
& +\sum_{K\in\mathds{Z}+\tfrac{1}{2}}^M 2\left|\frac{\varrho_{10}}{(K+\tfrac{1}{2} )!}- \frac{\varrho_{21}}{\sqrt{2}(K-\tfrac{3}{2})!}\right|^2.
\end{align} 

\begin{figure}
    \centering
    \includegraphics{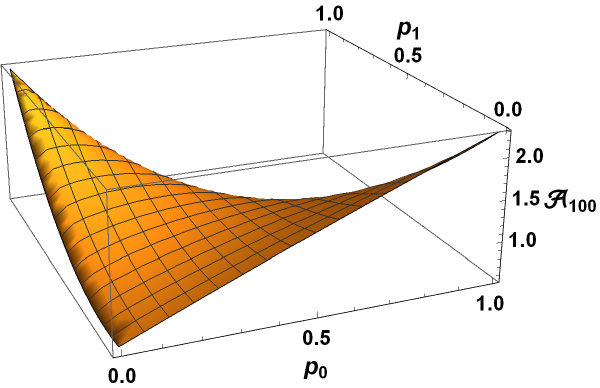}
    \caption{Cumulative multipole sum for optimal state $\ket{\psi_{\mathrm{opt}}}$ as a function of the two independent probabilities $p_0$ and $p_1$. The multipoles to order $M=100$ are included, by which point they have converged well beyond machine precision. It is clear that the maximum is obtained by setting all of the probability to go to either $p_0$ or $p_1$ with no shared probability between the two.}
    \label{fig:cum. multipoles vs p0 and p1}
\end{figure}

The relative phases that maximize this sum satisfy $2\psi-\phi=\pi$, so we can set $\eu^{\iu\psi}=1$ and $\eu^{\iu\phi}=-1$ without loss of generality. There are now only two constants to optimize over in the sum
\begin{align}
\mathfrak{A}_M(\ket{\psi_{\mathrm{opt}}})& =\sum_{K\in\mathds{Z}}^M\left|\frac{p_0}{K!}-\frac{p_1}{(K-1)!}+\frac{p_2}{2!(K-2)!}\right|^2+\frac{p_0p_2}{(K-1)!^2} \nonumber \\
& +\sum_{K\in\mathds{Z}+\tfrac{1}{2}}^M 2\left|\frac{\sqrt{p_0 p_1}}{(K+\tfrac{1}{2)}!}+\frac{\sqrt{p_1 p_2}}{\sqrt{2}(K-\tfrac{3}{2})!}\right|^2.
\end{align} 
All of the terms decay at least exponentially with $K$, so it is again evident that optimizing the sum for moderate $M$ will approximately optimize the sum for all larger $M$. Computing the contributions to $\mathfrak{A}_M$, we find 
\begin{align}
\mathfrak{A}_M(\ket{\psi_{\mathrm{opt}}})& \approx     
2.27959 p_0^2+2.27959 p_1^2+0.569896 p_2^2 \nonumber \\
& -0.622103 p_0 p_1+2.96853 p_0 p_2 +0.688948 p_1 p_2+1.94864 p_1 \sqrt{p_0 p_2},
\end{align}
which converges to this value by $M=7$ (see Fig. \ref{fig:coefficients converging}) and we have verified that these digits remain unchanged beyond $M=100$. This means that the sum will be maximized by either $p_0=1$ or $p_1=1$ (visualization in Fig. \ref{fig:cum. multipoles vs p0 and p1}). We can directly compute $\mathfrak{A}_M(\ket{0})-\mathfrak{A}_M(\ket{1})=1/\lfloor M\rfloor !^2$, where $\lfloor x \rfloor$ is the floor function that gives  the greatest integer less than or equal to $x$. This means that the vacuum state is the unique state with the maximal cumulative multipole moment for all $M$, while its supremacy diminishes exponentially with $M$.


\end{document}